\documentclass[twocolumn]{aastex7}

\usepackage{caption}
\usepackage{subcaption}

\begin{document}

\title{Detection of giant pulses from the transitional millisecond pulsar J1227$-$4853}

\author[orcid=0009-0005-1766-8992,gname=Saptarshi, sname=Sarkar]{Saptarshi Sarkar} 
\affiliation{National Centre for Radio Astrophysics, Tata Institute of Fundamental Research, Pune 411007, India}
\email{sarkar@ncra.tifr.res.in}

\author[orcid=0000-0002-2892-8025,gname=Jayanta, sname=Roy]{Jayanta Roy} 
\affiliation{National Centre for Radio Astrophysics, Tata Institute of Fundamental Research, Pune 411007, India}
\email{jroy@ncra.tifr.res.in}

\author[orcid=0000-0002-6287-6900,gname=Bhaswati, sname=Bhattacharyya]{Bhaswati Bhattacharyya} 
\affiliation{National Centre for Radio Astrophysics, Tata Institute of Fundamental Research, Pune 411007, India}
\email{bhaswati@ncra.tifr.res.in}

\author[orcid=0000-0002-5297-5278,gname=Paul, sname=Ray]{Paul S. Ray} 
\affiliation{Space Science Division, U.S. Naval Research Laboratory, Washington, DC 20375, USA}
\email{paul.ray@nrl.navy.mil}

\author[orcid=0000-0002-6631-1077,gname=Sanjay, sname=Kudale]{Sanjay Kudale} 
\affiliation{National Centre for Radio Astrophysics, Tata Institute of Fundamental Research, Pune 411007, India}
\email{ksanjay@ncra.tifr.res.in}

\author[orcid=0000-0003-1307-9435,gname=Paulo, sname=Freire]{Paulo Freire} 
\affiliation{Max-Planck-Institut f\"ur Radioastronomie, auf dem H\"ugel 69, D-53121 Bonn, Germany}
\email{pfreire@mpifr-bonn.mpg.de}

\author[orcid=0000-0003-2122-4540,gname=Patrick, sname=Weltevrede]{Patrick Weltevrede} 
\affiliation{Jodrell Bank Centre for Astrophysics, Department of Physics and Astronomy, University of Manchester, Manchester M13 9PL, UK}
\email{Patrick.Weltevrede@manchester.ac.uk}

\author[orcid=0000-0002-0893-4073,gname=Matthew, sname=Kerr]{Matthew Kerr} 
\affiliation{Space Science Division, U.S. Naval Research Laboratory, Washington, DC 20375, USA}
\email{matthew.kerr@nrl.navy.mil}

\correspondingauthor{Saptarshi Sarkar}
\email{sarkar@ncra.tifr.res.in}

\begin{abstract}

We report the discovery of giant pulse (GP) emission from the transitional millisecond pulsar (tMSP) PSR~J1227$-$4853, using 174 hours of single-pulse data from the upgraded Giant Metrewave Radio Telescope (uGMRT). This marks the first detection of GPs from a transitional MSP and adds to the small number of millisecond pulsars known to exhibit such extreme variability. A total of 235 GPs were detected across observations at 550--750~MHz, with widths as narrow as 1.28~$\mu$s and flux densities up to $\sim 10^4$ times the pulsar's mean flux density. The GPs are strongly localized in pulse phase, originating predominantly from the second and third main-pulse components, and are absent in the inter-pulse region. Their cumulative fluence distribution follows a power law above the completeness threshold, consistent with a defining characteristic of GP emission. The arrival times of the GPs deviate significantly from Poisson statistics, with the waiting-time distribution well described by a Weibull model having a shape parameter of $k = 0.30$, indicative of strong temporal clustering. During an epoch of enhanced activity, the GP rate increased by nearly two orders of magnitude to $124~\mathrm{hr}^{-1}$, with a corresponding shape parameter of $k = 0.47$. This value is similar to that reported for a burst storm from the repeating fast radio burst FRB~20200120E, suggesting possible phenomenological parallels between GPs from compact binary systems and repeating FRBs.

\end{abstract}


\section{Introduction} \label{sec:intro}

Millisecond pulsars (MSPs) are rapidly rotating neutron stars with spin periods $\lesssim 30$ ms, believed to have been spun up through long-term accretion of mass and angular momentum from a companion in a low-mass X-ray binary (LMXB) system~\citep{bhattacharya_formation_1991}. While this evolutionary link is well-established, only a handful of systems---PSR~J1023$+$0038~\citep{archibald_radio_2009,patruno_new_2013,takata_multi-wavelength_2014,stappers_state_2014}, PSR~J1824$-$2452I~\citep{papitto_swings_2013}, and PSR~J1227$-$4853~\citep{roy_discovery_2015}---have been observed to transition between accretion-powered (LMXB) and rotation-powered (MSP) states.

\begin{figure}
    \centering
    \includegraphics[width=\linewidth]{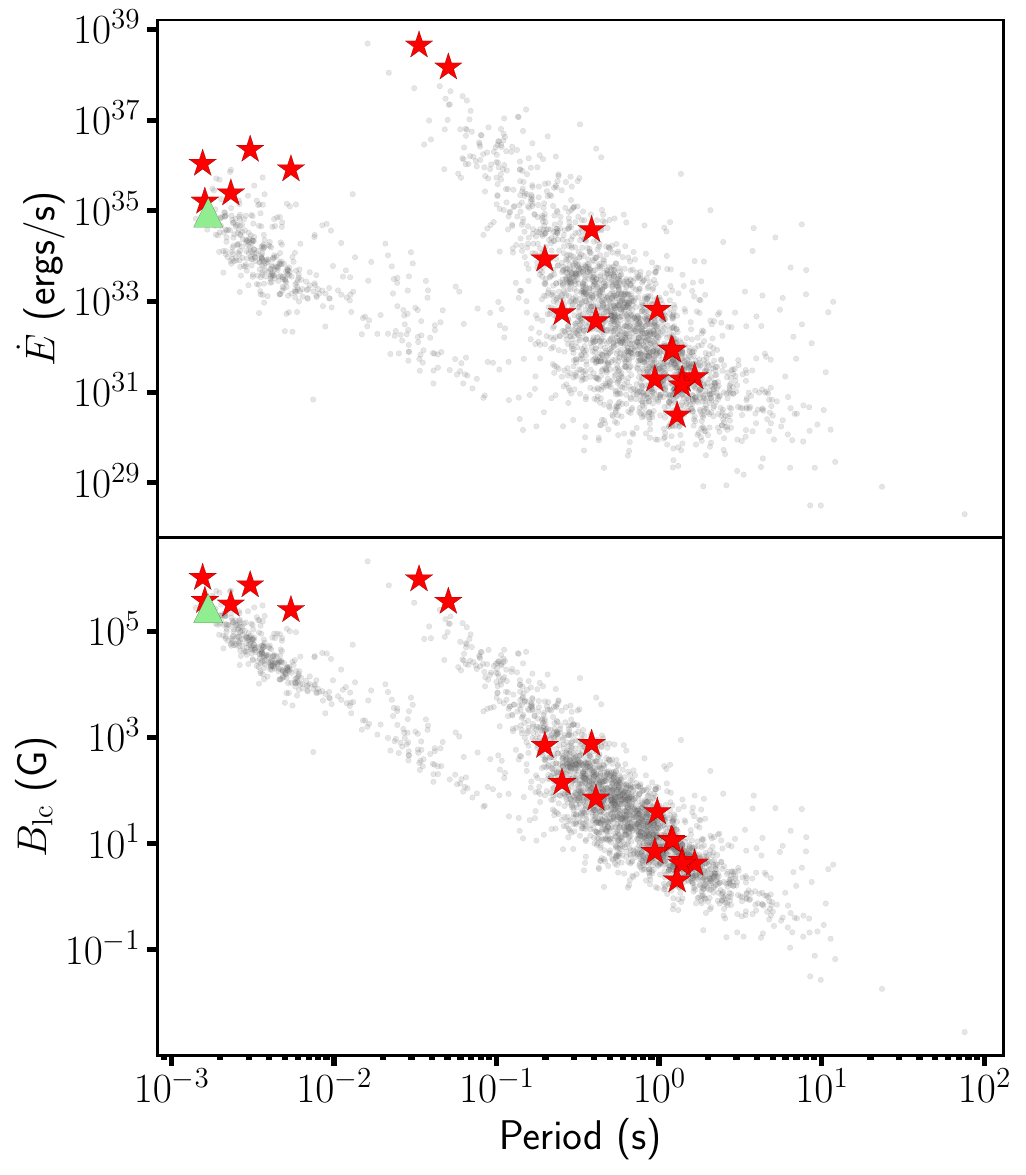}
    \caption{$B_{\rm lc}$ and $\dot{E}$ as a function of pulsar period for pulsars in the ATNF pulsar catalog (version 2.6.1). 19 pulsars reported to emit GPs are indicated by red stars; PSR~J1227$-$4853 is highlighted with a green triangle. The GP-emitting pulsars occupy three apparent groupings: recycled MSPs with high $B_{\rm lc}$, young pulsars with high $B_{\rm lc}$, and pulsars with comparatively low $B_{\rm lc}$ (and $\dot{E}$).}
    \label{fig:combined_scatter}
\end{figure}

The third such transitional MSP system, PSR~J1227$-$4853, was discovered using the Giant Metrewave Radio Telescope (GMRT) in 2014~\citep{roy_discovery_2015}. It is a 1.69 ms period eclipsing redback MSP~\citep{robertsSurroundedSpidersNew2012}. Figure \ref{fig:combined_scatter} presents the light-cylinder magnetic field strengths ($B_{\rm lc}$) and spin-down luminosities ($\dot{E}$) of pulsars from the ATNF catalog\footnote{\url{https://www.atnf.csiro.au/research/pulsar/psrcat}}~\citep{2005AJ....129.1993M}, showing that PSR~J1227$-$4853 ranks among the highest in both parameters relative to other known MSPs. Motivated by its high $B_{\rm lc}$ and $\dot{E}$, we investigate whether PSR~J1227$-$4853 exhibits giant pulse (GP) emission, which refers to extremely short-duration, high-intensity radio pulses reported from only 19 known pulsars to date~\citep{sun_detection_2021, malov_giant_2022}.

This is particularly compelling given that all known GP-emitting recycled MSPs have high values of $B_{\rm lc}$ ($>10^5$ G) (Figure \ref{fig:combined_scatter}), suggesting that the GP emission mechanism for MSPs may be linked to extreme conditions near the light cylinder~\citep{cognard_giant_1996}. However, with only 5 GP-emitting MSPs reported to date~\citep{malov_giant_2022}, the physical drivers of this phenomenon remain poorly understood. Increasing the number of such known sources is therefore crucial for constraining theoretical models of GP emission mechanisms.

In this paper, we present results from a single-pulse analysis of upgraded GMRT (uGMRT)~\citep{swarup_giant_1991,gupta_upgraded_2017} observations of PSR~J1227$-$4853---the first such search reported for a transitional MSP. The rest of the paper is organized as follows: Section \ref{sec:observations} describes the observations; Section \ref{sec:analysis} details the analysis procedures; Section \ref{sec:results} presents our findings; and Section \ref{sec:discussion} explores the implications of the detected pulses. We summarize our conclusions in Section \ref{sec:summary}.

\section{Observations} \label{sec:observations}

This study utilizes data from regular timing observations of the transitional MSP PSR~J1227$-$4853 with the uGMRT. Observations were conducted using the band 4 receiver system with a bandwidth of 200 MHz (550--750 MHz) in phased array mode, with coherent dedispersion applied to remove intra-channel dispersion delays.

Two distinct observing setups were employed, yielding two datasets with differing time-resolutions:

\begin{itemize}
    \item The first dataset comprises 165 hours of observations across 185 epochs between 2019 January 13 and 2024 August 9. These data were recorded with the 200 MHz bandwidth subdivided into 512 spectral channels, coherently dedispersed and sampled at 10.24~$\mu$s.
    
    \item The second dataset comprising 9 hours of observations was acquired from more recent observations across 10 epochs between 2024 September 8 and 2025 February 16. These data were recorded using the same 200 MHz bandwidth, but subdivided into 64 spectral channels, coherently dedispersed and with finer time sampling at 1.28~$\mu$s. These data were collected to enable detection of extremely narrow GPs potentially missed in earlier observations.
\end{itemize}

\section{Data Analysis} \label{sec:analysis}

\subsection{Single pulse search} \label{subsec:singlepulsesearch}

For the 10.24~$\mu$s resolution data, the coherently dedispersed raw telescope data, consisting of a header-less dynamic spectrum at the native time and frequency resolution provided by the beamformer, were first processed using the GMRT Pulsar Tool (\texttt{gptool}\footnote{\url{https://github.com/chowdhuryaditya/gptool}}) to mitigate radio frequency interference (RFI). As part of this processing, \texttt{gptool} also performs bandpass normalization. Filterbank files, at the required time and frequency resolution for
subsequent processing and analysis, were created using \texttt{SIGPROC}'s \texttt{filterbank} program~\citep{lorimerSIGPROCPulsarSignal2011}, and incoherently dedispersed at PSR~J1227$-$4853's nominal dispersion measure (DM) of 43.449 pc cm$^{-3}$ using \texttt{PRESTO}'s \texttt{prepdata} 
routine~\citep{2011ascl.soft07017R} to remove inter-channel dispersion smearing. The dedispersed time series were searched for single pulses using \texttt{PRESTO}'s \texttt{single\_pulse\_search.py} script, using a signal-to-noise ratio (S/N) threshold of 7. This threshold was chosen as it led to the detection of a large number of GP candidates, while keeping the number of spurious candidates low. The \texttt{single\_pulse\_search.py} script performs matched filtering using boxcar templates of increasing widths. We used boxcars with widths 1, 2, 3, 4, 6, 9, 14, and 20 bins (i.e. over a pulse width range of 10.24~$\mu$s to 204.8~$\mu$s). The boxcar width maximizing the S/N was adopted as a proxy for the pulse width; parametric profile fitting was not attempted because several GPs span only a few time bins, making parametric width estimates poorly constrained. Finally, candidates exhibiting clear signatures of RFI in their dedispersed dynamic spectra were rejected. To identify likely RFI, candidates showing narrow-band emission confined to only a few frequency channels (typically one or two channels) or exhibiting temporally extended structures were discarded. Running \texttt{single\_pulse\_search.py} on the RFI-mitigated data yielded 1510 candidates, of which 1287 ($\sim85\%$) were rejected following visual inspection.

For the 1.28~$\mu$s time-resolution data, automated RFI mitigation was not feasible due to incompatibility of \texttt{gptool} with such high-time-resolution data. Consequently, the raw telescope data were converted directly to filterbank format using \texttt{DSPSR}'s \texttt{digifil} utility~\citep{stratenDSPSRDigitalSignal2011}, which also performs bandpass normalization. For high-time-resolution data, even a small error in the DM can cause residual smearing of the signal, reducing its S/N below the detection threshold. For example, given our 200 MHz bandwidth, a DM error of 0.005 pc cm$^{-3}$ introduces a smearing of $\sim 30~\mu$s --- comparable to our maximum bin size of $25.6~\mu$s. While this is not a significant concern for the 10.24~$\mu$s resolution data (with a maximum bin size of 204.8~$\mu$s), it significantly impacts the 1.28~$\mu$s resolution data, leading to non-detection of several narrow, moderate S/N GPs. To mitigate this, we employed a DM optimization procedure during the GP search for the high-time-resolution data. We generated dedispersed time series for 20 different DMs centered on the pulsar's nominal DM, in steps of 0.001 pc cm$^{-3}$, using \texttt{PRESTO}'s \texttt{prepsubband} routine. To prevent any residual intra-subband smearing from the algorithm's typical tiered approach, we set the number of subbands equal to the total number of frequency channels of the data. Each dedispersed time series was searched for single pulses using the same matched filtering approach and the DM that led to the maximum S/N for each candidate was recorded along with the corresponding width. Since automated RFI filtering was unavailable, each candidate's dedispersed dynamic spectrum was manually inspected for signs of RFI. Running \texttt{single\_pulse\_search.py} on the unprocessed data yielded 124 candidates, of which 112 ($\sim90\%$) were identified as RFI.

\subsection{S/N optimization} \label{subsec:optimization}

In both datasets, the DM of each candidate pulse was refined to maximize S/N. For the 10.24~$\mu$s resolution data, this was done by using \texttt{prepdata} to dedisperse each candidate single pulse at 20 different DMs centered on the pulsar's nominal DM, in steps of 0.001 pc cm$^{-3}$, and listing the DM that led to the highest S/N. For each candidate, the optimized DM and associated pulse width were recorded. For the 1.28~$\mu$s resolution data, this optimization was intrinsically built into the search pipeline as described in Section \ref{subsec:singlepulsesearch}. 

Finally, the single pulses with optimized pulse widths $< 150~\mu$s were saved for further analysis. A cutoff width of $150~\mu$s was selected as it corresponds to the typical width of the main-pulse component in the average pulse profile of PSR~J1227$-$4853. Since GPs are expected to be narrower than the components of a pulsar's average profile, this threshold serves as a conservative upper limit. Additionally, coupled with the S/N threshold of 7, it corresponds to a flux density detection threshold of 0.4~Jy---300 times greater than the mean flux density of PSR~J1227$-$4853 at 650 MHz, which is measured to be 1.2~mJy~\citep{kudale_study_2020}. The fluence threshold depends on the minimum searched pulse width allowed by our time resolution, with the $1.28~\mu$s data corresponding to a fluence detection threshold of $3.8~\mathrm{Jy}\,\mu\mathrm{s}$.

\section{Results} \label{sec:results}

\subsection{Giant Pulses} \label{subsec:giantpulses}

\begin{figure}
    \centering
    \includegraphics[width=\linewidth]{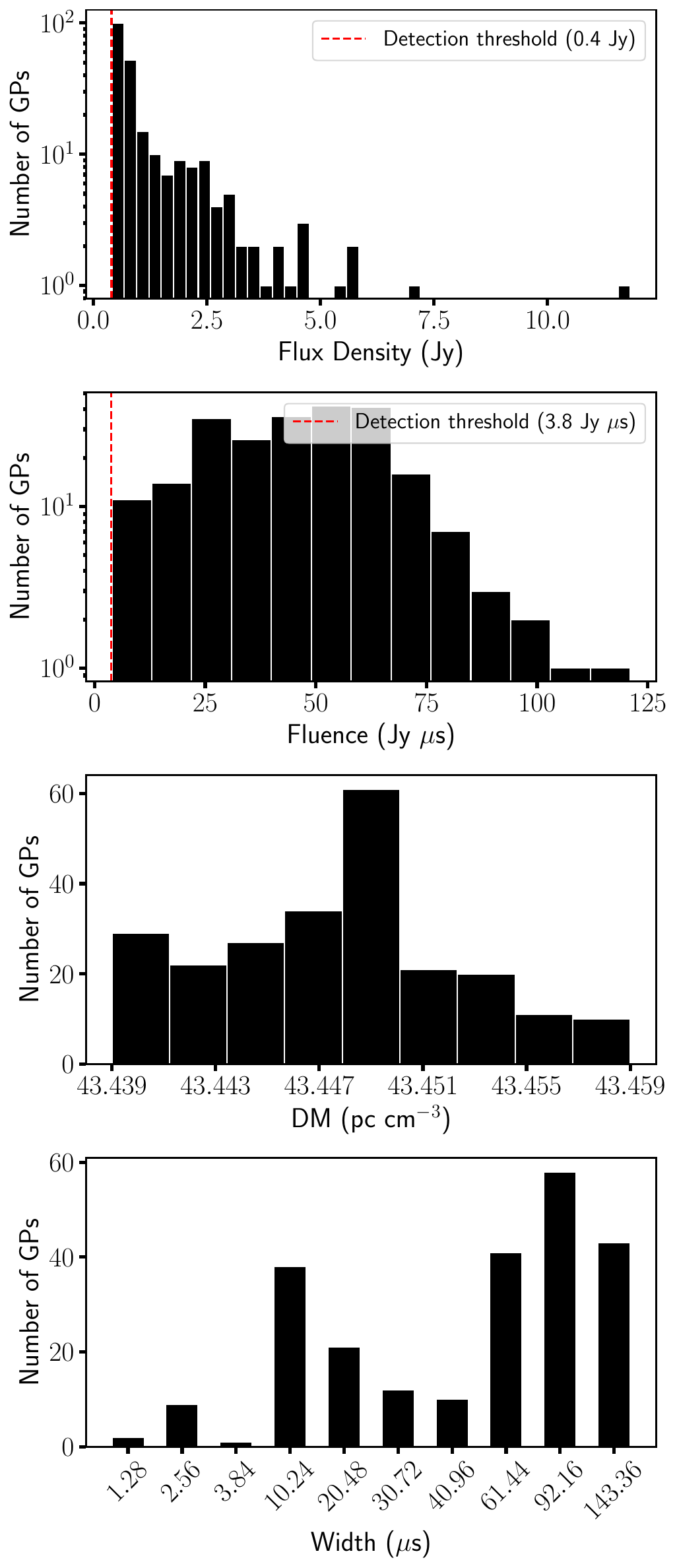}
    \caption{Distribution of flux densities, fluences, DMs, and widths of the GPs. Note that the width axis is not linearly scaled.}
    \label{fig:histograms}
\end{figure}

\begin{figure}
    \centering
    \includegraphics[width=\linewidth]{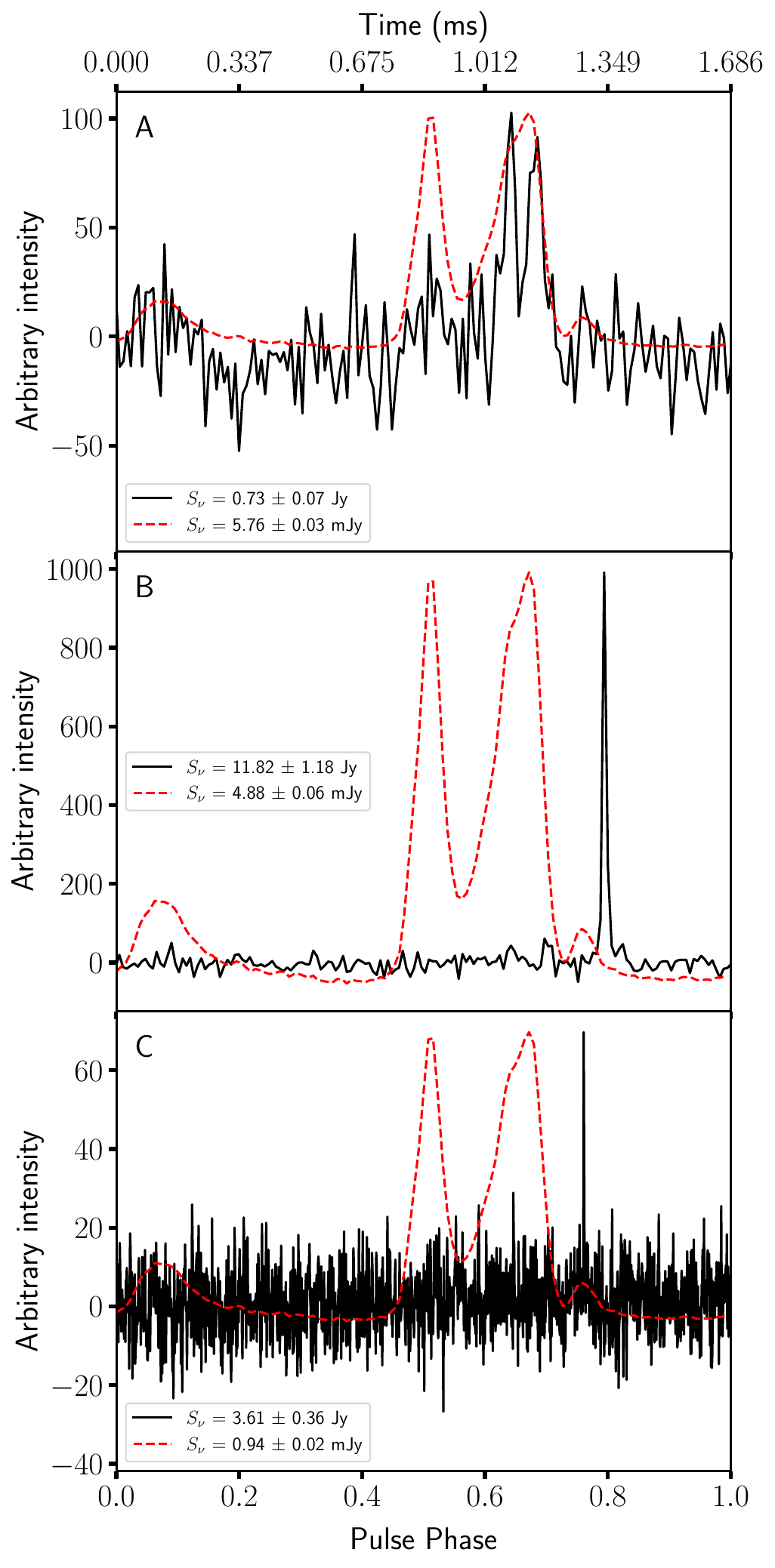}
    \caption{Three examples of GPs detected during our search. The average profile of PSR~J1227$-$4853 is overlaid in red. The mean flux density during the observation epoch, and the flux density of the GP (calculated using the radiometer equation) are also labeled.
    \textit{Panel A:} One of the broadest GPs (S/N = 12.54, Width = 143.36~$\mu$s). 
    \textit{Panel B:} The brightest GP (S/N = 54.98, Width = 10.24~$\mu$s). 
    \textit{Panel C:} One of the narrowest GPs (S/N = 8.57, Width = 1.28~$\mu$s).}
    \label{fig:example_Gps}
\end{figure}

\begin{figure*}
     \centering
     \begin{subfigure}[b]{0.45\textwidth}
         \centering
         \includegraphics[width=\textwidth]{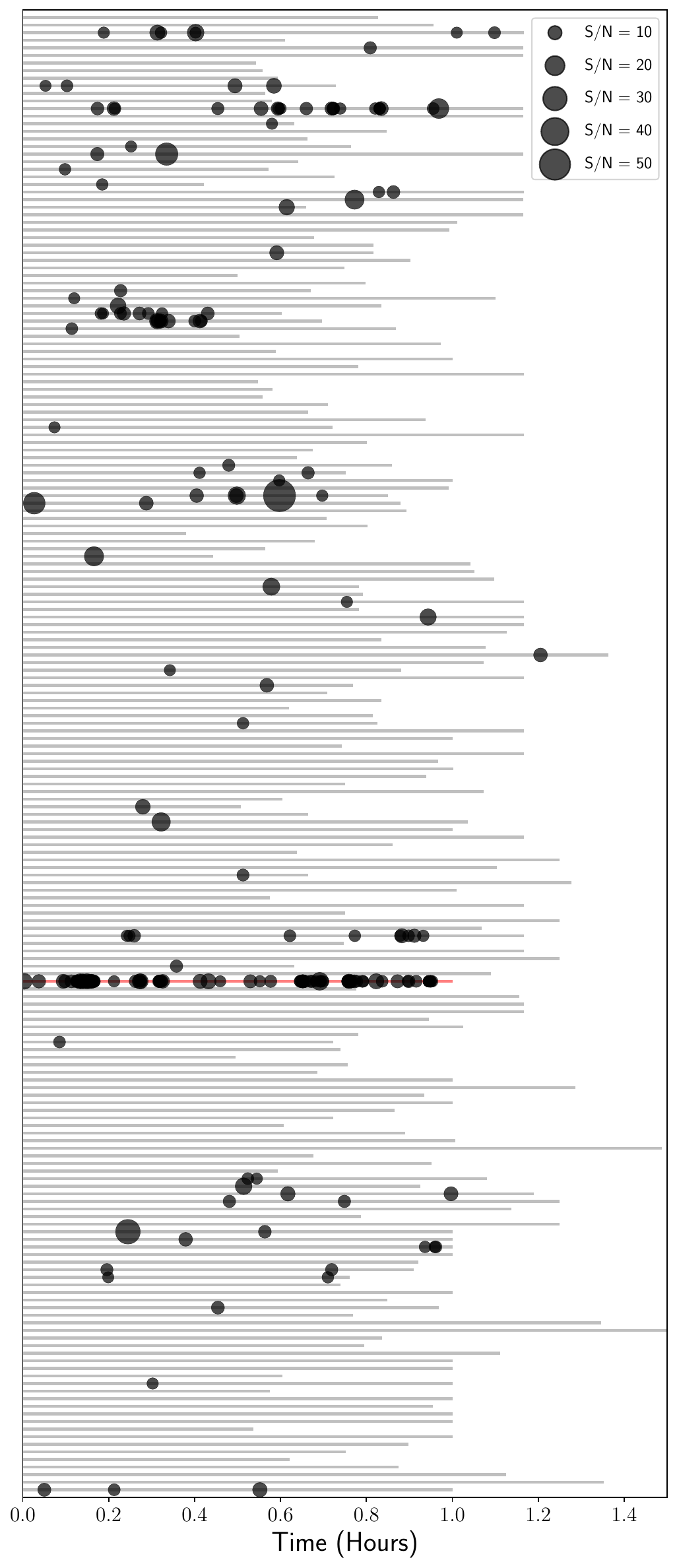}
     \end{subfigure}
     \hspace{0.5cm}
     \begin{subfigure}[b]{0.45\textwidth}
         \centering
         \includegraphics[width=\textwidth]{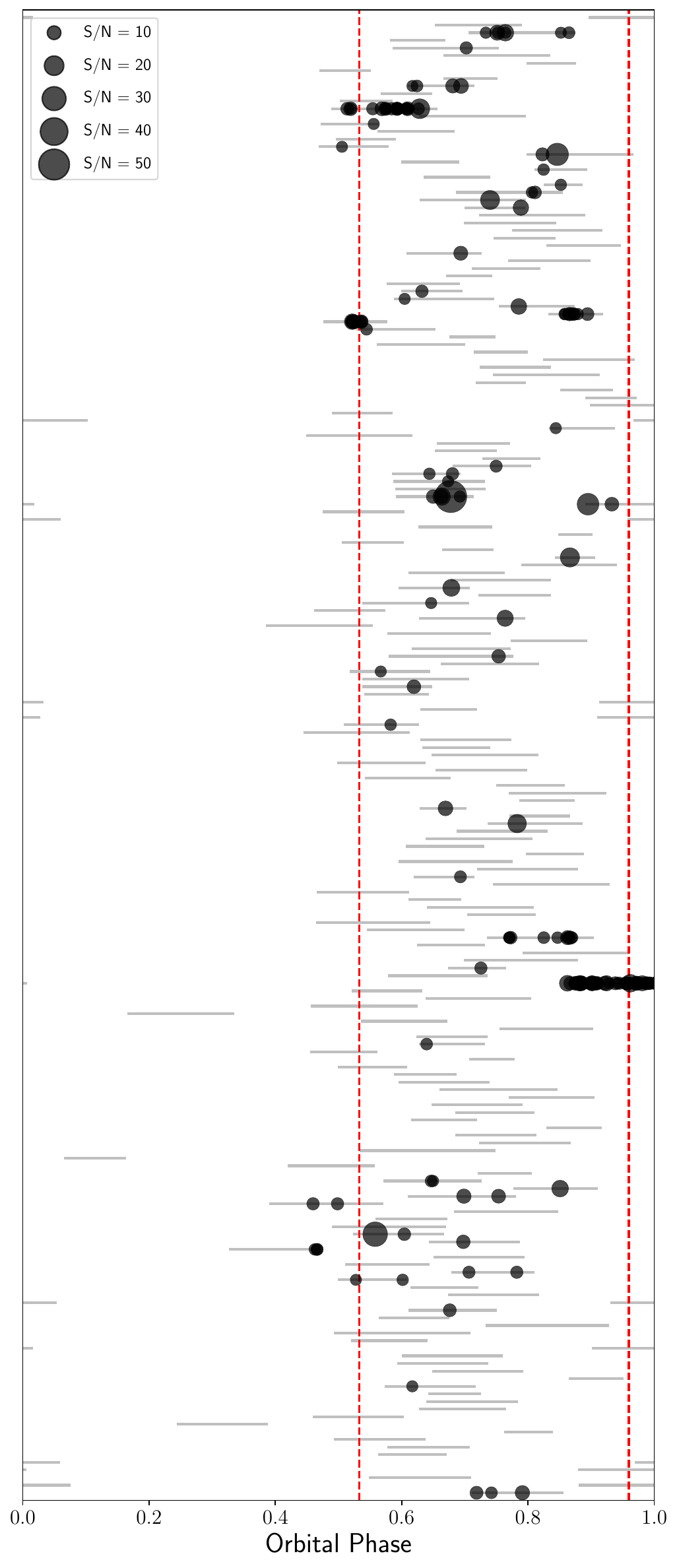}
     \end{subfigure}
    \caption{Observation intervals (grey lines) with the positions of detected GPs overlaid, where marker size indicates S/N. \textit{Left panel:} Observations aligned to a common start time and ordered chronologically, with the most recent observation at the top; the 2020 December 6 epoch is highlighted in red. \textit{Right panel:} The same observations shown in orbital phase, with red dashed lines marking the eclipse boundaries for band~4~\citep{kudale_study_2020}; the superior conjunction occurs at orbital phase 0.25.}
     \label{fig:observations_combined}
     \hfill
\end{figure*}

\begin{figure}
    \centering
    \includegraphics[width=\linewidth]{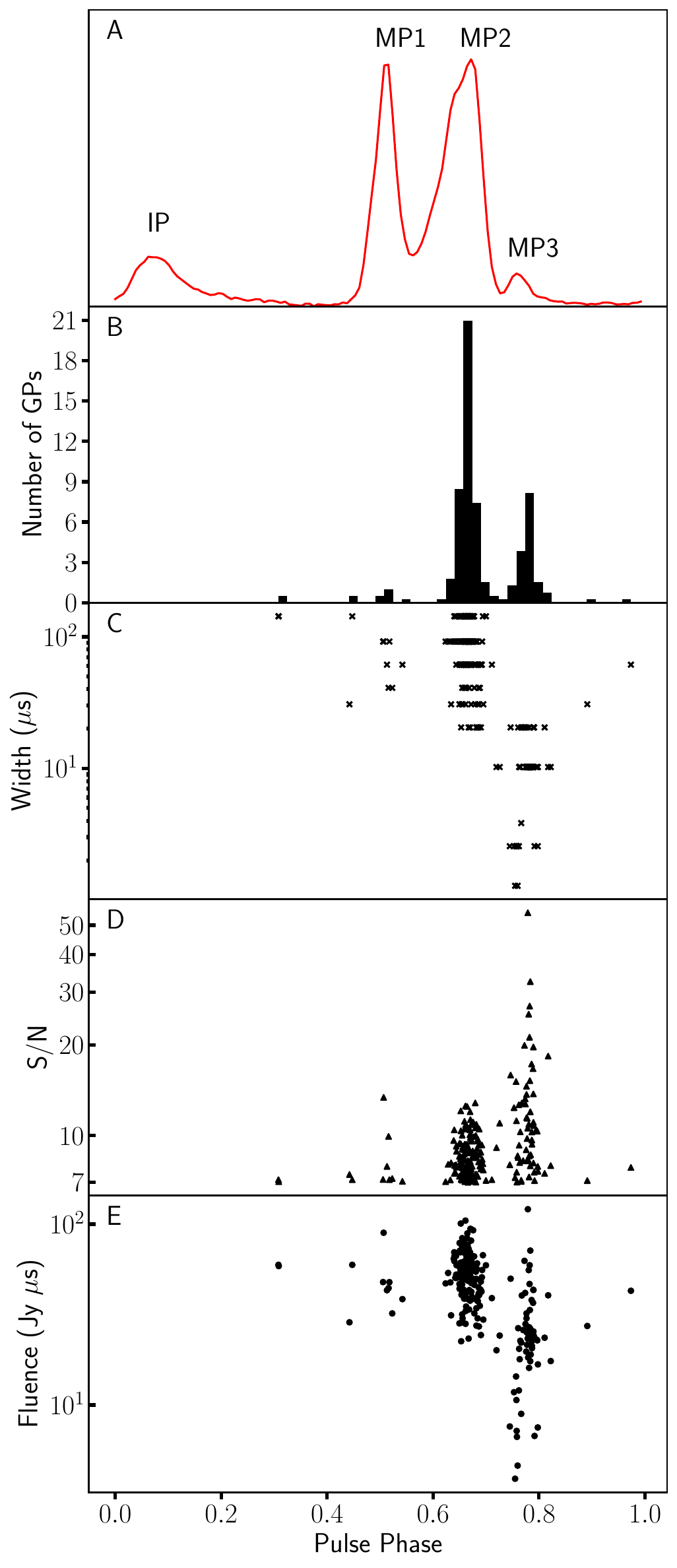}
    \caption{\textit{Panel A:} Average pulse profile for PSR~J1227$-$4853 (in arbitrary units). The inter-pulse (IP) and the three main-pulse components are marked (MP1, MP2, and MP3). \textit{Panel B:} Distribution of the number of GPs detected at different pulse phases. \textit{Panel C:} Distribution of GP widths across pulse phases. \textit{Panel D:} Distribution of GP S/N across pulse phases.  \textit{Panel E:} Distribution of GP fluences across pulse phases.}
    \label{fig:pp_combined}
\end{figure}

A total of 235 GPs were detected as a result of our search. Figure \ref{fig:histograms} shows the distribution of flux densities and fluences of the detected GPs, computed using the GMRT exposure time calculator\footnote{\url{https://www.gmrt.ncra.tifr.res.in/gtac/etc/etc/rmsp_advanced/rmsp.html}} based on the radiometer equation using the optimized pulse widths and S/N, along with the distributions of their DMs and widths. Figure \ref{fig:example_Gps} shows examples of three detected GPs with varying pulse widths ranging from 143.36~$\mu$s to 1.28~$\mu$s. Figure~\ref{fig:observations_combined} shows the observation intervals and the locations of the detected GPs as a function of time (left panel) and orbital phase (right panel). Notably, a significant increase in the GP rate occurred on 2020 December 6 (MJD 59189), with 114 GPs being detected within a single one-hour observation on that date. The distribution of GPs across pulse phase is displayed in Figure \ref{fig:pp_combined}B. We detected no GPs in the inter-pulse region, with the majority of pulses concentrated in the phases corresponding to the main-pulse components MP2 and MP3. The pulse width distribution as a function of pulse phase is presented in Figure \ref{fig:pp_combined}C. 

Figures \ref{fig:example_Gps} and \ref{fig:pp_combined} indicate that the properties of the detected GPs depend strongly on the pulse-phase region from which they originate. GPs associated with the MP3 component are systematically narrower than those associated with MP2. As shown in \ref{fig:pp_combined}C, all MP3-associated GPs have widths $\leq 20.48~\mu$s, with the narrowest detected pulses (down to 1.28~$\mu$s) originating exclusively from this phase window. In addition, Figure \ref{fig:pp_combined}D shows that the highest S/N events are predominantly concentrated in the MP3 phase region. We also show the distribution of fluences in Figure \ref{fig:pp_combined}E.

\subsection{Fluence distribution} \label{subsec:fluencedist}

\begin{figure}
    \centering
    \includegraphics[width=\linewidth]{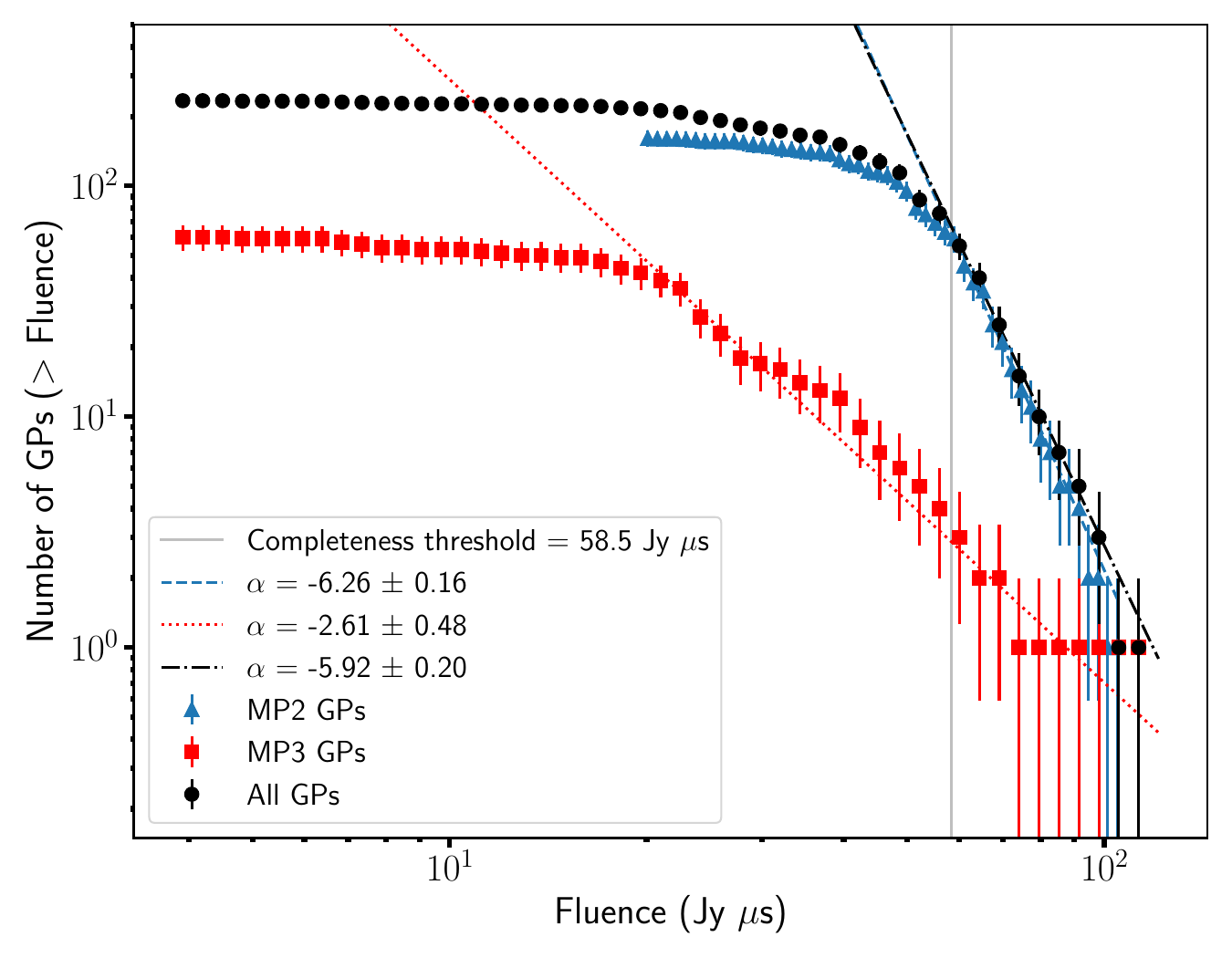}
    \caption{Cumulative fluence distributions of the detected giant pulses from the MP2 (blue triangles), MP3 (red squares), and combined (black circles) samples. The error bars correspond to Poissonian errors. The vertical grey line marks the fluence completeness threshold. The dashed lines show the best-fit power-law models fitted above the completeness threshold.}
    \label{fig:fluence_cdfs}
\end{figure}

One of the key defining characteristics of GPs is a power-law pulse energy (fluence) distribution. Figure~\ref{fig:fluence_cdfs} shows the cumulative fluence distributions for the MP2, MP3, and combined GP samples. We fit power-law models above the fluence completeness threshold, defined following the prescription of~\citet{keane_fast_2015} as the fluence corresponding to a pulse with $\mathrm{S/N}=7$ and the maximum pulse width of $143.36~\mu$s in our dataset. This criterion ensures that pulses above the threshold are detectable over the full range of pulse widths present in the sample.

The cumulative distributions above this threshold were fitted with a power-law model of the form
\begin{equation}
N(>F) = A F^{\alpha},
\end{equation}
where $F$ is the pulse fluence, $A$ is a normalization constant, and $\alpha$ is the cumulative power-law index. The best-fit values are $\alpha = -5.92$ for the combined sample, $\alpha = -6.26$ for MP2, and $\alpha = -2.61$ for MP3. The best-fit models are shown in Figure~\ref{fig:fluence_cdfs}.

\subsection{Waiting Times} \label{subsec:waitingtimes}

If GPs arise from a Poisson process, the distribution of waiting times (i.e., the time intervals $\delta$ between successive GPs) should follow an exponential form:

\begin{equation} P(\delta \mid r) = r e^{-r \delta}, \end{equation}

where $r$ is the GP rate, and $P(\delta \mid r)$ is the conditional probability density of $\delta$ given $r$.

In the case of the Crab pulsar, GPs are generally found to exhibit an exponential waiting-time distribution~\citep{lundgren_giant_1995,karuppusamy_giant_2010}. Similar behavior has also been reported for GP emission from several other pulsars, suggesting that approximately Poissonian pulse arrival statistics may be a common feature of at least a subset of GP-emitting neutron stars~\citep[e.g.,][]{mckeeDetailedStudyGiant2019,hoDetection370002025,vanruitenSearchingLinksEnergetic2026}. Exponential waiting-time distributions have likewise been observed in several RRATs, indicating comparable pulse-arrival statistics in some sporadically emitting neutron-star sources~\citep[e.g.,][]{chenDiscoveryRotatingRadio2022,xieEmissionPropertiesRRAT2022,hsuExploringSinglepulseBehaviours2023,zhongStudyingRadiationCharacteristics2024}. However, certain single-pulse sources, such as FRB 20121102A, exhibit significant deviations from this behavior. For FRB 20121102A, the waiting-time distribution is better described by the Weibull distribution~\citep{oppermann_non-poissonian_2018}

\begin{equation} \mathcal{W}(\delta \mid k, r) = k \delta^{-1} \left[ \delta r \Gamma(1 + 1/k) \right]^k e^{- [\delta r \Gamma(1 + 1/k)]^k}, 
\end{equation}

where $k$ is the shape parameter and $\Gamma(x)$ is the gamma function. The shape parameter characterizes the degree of deviation from Poissonian behavior: $k = 1$ corresponds to a pure exponential distribution (i.e., a Poisson process), $k < 1$ indicates temporal clustering, and the limit $k \to \infty$ corresponds to periodic emission.

\begin{figure*}
     \centering
     \begin{subfigure}[b]{0.45\textwidth}
         \centering
         \includegraphics[width=\textwidth]{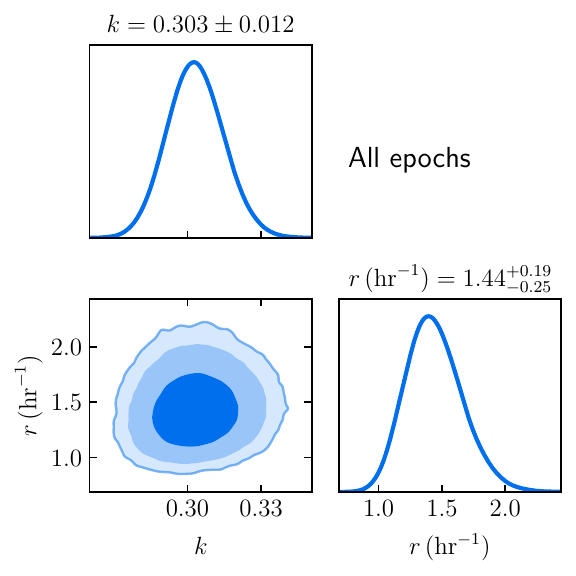}
     \end{subfigure}
     % \hfill
     \hspace{0.5cm}
     \begin{subfigure}[b]{0.45\textwidth}
         \centering
         \includegraphics[width=\textwidth]{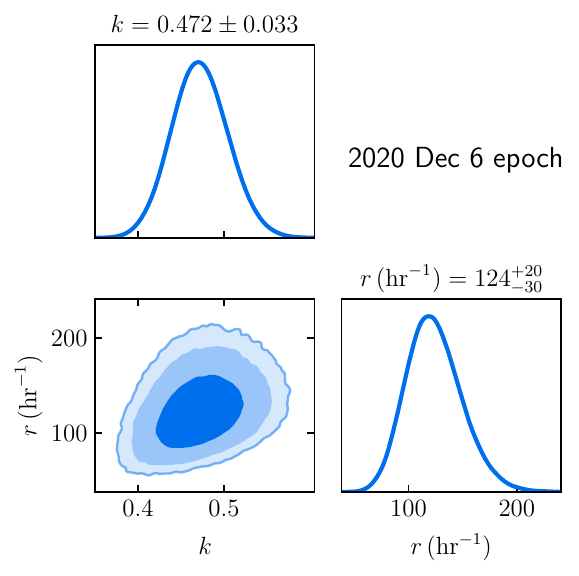}
     \end{subfigure}
     \caption{Posterior distribution for the two Weibull parameters $k$ and $r$ for all epochs combined (left panel) and for the 2020 December 6 epoch (right panel). The contours correspond to the 68 per cent, 95 per cent, and 99 per cent confidence intervals. The mean and 68 per cent confidence intervals are denoted above the one-dimensional posterior of the respective parameters.}
     \label{fig:waiting_times}
     \hfill
\end{figure*}

To investigate potential clustering in the arrival times of GPs from PSR~J1227$-$4853, we follow the method of~\citet{oppermann_non-poissonian_2018} to obtain posterior distributions for the Weibull parameters $k$ and $r$. Assuming independent epochs, we compute the total likelihood $\mathcal{L}(\mathcal{D} \mid k, r)$ as the product of the individual likelihoods for each observation. We adopt a uniform prior $\mathcal{P}(k, r) = \text{const.}$, and use the Markov chain Monte Carlo (MCMC) sampler package \texttt{emcee}~\citep{foreman-mackey_emcee_2013} to explore the posterior distribution

\begin{equation}
    \mathcal{P}(k, r \mid \mathcal{D}) \propto \mathcal{L}(\mathcal{D} \mid k, r)\mathcal{P}(k, r).
\end{equation}

We also perform a similar analysis for the 2020 December 6 epoch separately, due to its elevated GP rate and large number of detected GPs. Figure \ref{fig:waiting_times} shows the posterior distributions for both cases. For the combined dataset, the posterior distribution yields a mean shape parameter of $k = 0.303 \pm 0.012$ and a mean GP rate of $r = 1.44^{+0.19}_{-0.25}$ hr$^{-1}$, with uncertainties corresponding to the 68 per cent confidence intervals. For the 2020 December 6 epoch alone, the posteriors yield $k = 0.472 \pm 0.033$ and $r = 124^{+20}_{-30}$ hr$^{-1}$.

We note that the Weibull formalism used here provides only a phenomenological single-scale description of clustering and implicitly assumes a stationary process governing the burst arrival statistics. More complex behaviors, such as time-dependent activity states or multi-mode clustering, are not naturally captured within this framework. In addition, finite observing windows and incomplete sampling may bias the inferred clustering parameter toward stronger apparent clustering. Alternative stochastic-process approaches, such as Markov-modulated Poisson processes~\citep{fischerMarkovmodulatedPoissonProcess1993} or self-exciting Hawkes processes~\citep{bonnetMaximumLikelihoodEstimation2021}, may provide a more physically realistic description of burst activity, particularly for sources exhibiting state changes or correlated triggering. A detailed investigation of such models is beyond the scope of the present work, but we consider this to be an interesting direction for a future study.

\section{Discussion} \label{sec:discussion}
We detected no GPs at the pulse phase corresponding to the inter-pulse region, in contrast to GP-emitting pulsars such as the Crab~\citep{cordesBrightestPulsesUniverse2004,karuppusamy_giant_2010,karuppusamy_crab_2012,mickaligerGIANTSAMPLEGIANT2012,doskoch_statistical_2024}, PSR~B1937$+$21~\citep{knightParkesRadioTelescope2007,mckeeDetailedStudyGiant2019}, and PSR J1823$-$3021A~\citep{abbateGiantPulsesJ18233021A2020,hoDetection370002025}, where GPs are detected at both the main-pulse and inter-pulse phases. This suggests that in PSR~J1227$-$4853, GP emission is localized to a specific region in the magnetosphere associated with the main-pulse. Unlike their pulse-phase distribution, however, we find that the GPs show no significant dependence on orbital phase (see Figure~\ref{fig:observations_combined} (right panel)), in contrast to eclipsing binary systems such as PSR~B1957$+$20 where strong orbital-phase-dependent pulse magnification and variability arise from plasma lensing in the companion's outflow~\citep{main_pulsar_2018}. This lack of orbital-phase dependence argues against propagation effects or lensing in the binary environment as the primary driver of the observed GP properties. 

An important question raised by this discovery is why PSR~J1227$-$4853 is the first tMSP from which GP emission has been detected. At present, this may largely reflect observational limitations rather than an intrinsic property of the source. Of the three confirmed tMSPs, PSR~J1023$+$0038 has remained in an LMXB-like state since 2013~\citep{patruno_new_2013,stappers_state_2014}, during which radio pulsations are not observed, preventing comparable GP searches. The third system, PSR~J1824$-$2452I (M28I), has not been the subject of dedicated GP studies, although single-pulse searches in M28 have been conducted for the unrelated pulsar PSR~J1824$-$2452A~\citep{vanruitenSearchingLinksEnergetic2026}, which is itself a known GP emitter. Consequently, PSR~J1227$-$4853 currently represents the only tMSP that is both observable as a radio pulsar and has been extensively searched for GP emission. Furthermore, PSR~J1227$-$4853 was selected for this study because it possesses one of the highest light-cylinder magnetic field strengths among known MSPs, a property already associated with GP emission in recycled pulsars. Therefore, the present detection can be explained without invoking any mechanism unique to transitional MSPs.

The cumulative fluence distributions of the detected pulses are well described as power-laws above the fluence completeness threshold (Figure~\ref{fig:fluence_cdfs}), consistent with one of the defining observational characteristics of GP emission. Interestingly, the inferred power-law indices differ significantly between the MP2 and MP3 phase regions. The MP3-associated pulses exhibit a comparatively shallow slope ($\alpha=-2.61$), consistent with the canonical power-law indices commonly reported for GP-emitting pulsars such as the Crab pulsar~\citep{karuppusamy_giant_2010} and PSR~B1937$+$21~\citep{mckeeDetailedStudyGiant2019}. In contrast, the MP2-associated pulses exhibit a much steeper fluence distribution ($\alpha=-6.26$), although similarly steep indices have also been reported for individual observations of PSR~B1937$+$21~\citep{mckeeDetailedStudyGiant2019}.

Furthermore, all GPs with widths $\leq 10~\mu$s originate from a narrow pulse phase window corresponding to one of the main-pulse components, and these pulses also exhibit the highest S/N (see Figures~\ref{fig:pp_combined}C and \ref{fig:pp_combined}D). The clear separation in width and S/N distributions between the MP2 and MP3-associated pulses suggests that the two phase regions may correspond to distinct GP emission regimes. However, unlike the width and S/N distributions, the fluence distributions of the MP2 and MP3-associated pulses show a weaker distinction, with the MP3-associated pulses tending to exhibit lower fluences (see Figure~\ref{fig:pp_combined}E).

We also searched for any preferential ordering between consecutive MP2 and MP3-associated GPs within individual observing epochs. While consecutive pulses frequently originate from the same phase region, we do not find any evidence for a preferred ordering between the MP2 and MP3 populations, with the numbers of MP2$\rightarrow$MP3 and MP3$\rightarrow$MP2 transitions being identical within our sample.

The extremely narrow widths and high S/N values of the MP3-associated pulses are reminiscent of canonical GPs observed in sources such as the Crab pulsar, which are characterized by very short durations and high peak intensities. In contrast, the broader and generally lower S/N pulses detected from the MP2 phase region may represent the high-fluence tail of a broader single-pulse energy distribution, similar to the so-called `giant micropulses' observed in pulsars such as Vela~\citep{johnstonHighTimeResolution2001}. While both populations satisfy our operational GP selection criteria, their distinct observational properties point to potentially different emission conditions or magnetospheric regions within the same pulsar.

The waiting time analysis, combining all epochs, yields a Weibull shape parameter of $k = 0.303 \pm 0.012$ and a mean GP rate of $r = 1.44^{+0.19}_{-0.25}$ hr$^{-1}$. The sub-Poissonian value of $k$ indicates strong temporal clustering of GP arrival times, deviating from the Poissonian statistics observed in sources like the Crab pulsar~\citep{lundgren_giant_1995, karuppusamy_giant_2010}. During the enhanced activity on 2020 December 6, the GP rate increased significantly to $r = 124^{+20}_{-30}$ hr$^{-1}$, with a corresponding $k = 0.472 \pm 0.033$. Interestingly, this is consistent with $k=0.50^{+0.04}_{-0.05}$ measured from an FRB 20200120E burst storm~\citep{nimmo_burst_2023}. While this statistical similarity alone does not establish a physical connection, it raises the possibility that GPs from compact binary systems, such as PSR~J1227$-$4853, may exhibit phenomenology analogous to repeating FRBs. This hypothesis is further reinforced by the localization of FRB 20200120E to an old globular cluster in M81~\citep{kirsten_repeating_2022}, which poses a challenge to models invoking young magnetars and instead favors scenarios involving compact binary systems known to form efficiently in dense stellar environments like globular clusters.

\section{Summary} \label{sec:summary}

In this work, we have reported and characterized the first detection of GP emission from the transitional MSP PSR~J1227$-$4853. Our long-term uGMRT monitoring at 550--750 MHz has revealed a total of 235 GPs with flux densities up to $\sim 10^4$ times the pulsar's mean flux density and pulse widths as narrow as 1.28~$\mu$s, establishing this system as a new member of the rare class of GP-emitting pulsars.

The GPs are strongly phase-localized, predominantly originating from the second and third main-pulse components, and are absent in the inter-pulse region, indicating a highly localized magnetospheric origin. In contrast, their occurrence shows no significant dependence on orbital phase. Their cumulative fluence distributions are well described by power laws above the completeness threshold, further supporting their classification as GPs. The distribution of GP waiting times exhibits strong deviations from Poisson statistics, well-modeled by a Weibull distribution with a shape parameter $k = 0.303$, which indicates strong temporal clustering. During an epoch of elevated activity, the GP rate increased to $r = 124$ hr$^{-1}$, accompanied by a $k = 0.472$. This value is statistically consistent with that observed for repeating FRB 20200120E during a burst storm, suggesting potential phenomenological parallels between GPs from compact binaries and repeating FRBs.

As PSR~J1227$-$4853 is currently the only transitional MSP both observable in the radio pulsar state and extensively searched for GP emission, it remains unclear whether GP activity is a common property of transitional MSPs or instead reflects the extreme magnetospheric conditions of this source. Future single-pulse studies of other transitional systems will be required to distinguish between these possibilities. Further, the phenomenological parallels, particularly with the waiting-time statistics of FRB 20200120E, raise intriguing questions about the commonality of underlying physical processes. 

Overall, this discovery expands the phenomenological landscape of pulsar emission, and highlights the importance of systematic single-pulse studies in unveiling the diversity of radio emission mechanisms in compact binaries.

\begin{acknowledgments}
We acknowledge the support of the Department of Atomic Energy, Government of India, under project No. 12-R\&D-TFR-5.02-0700. We thank the staff of the GMRT that made these observations possible. GMRT is run by the National Centre for Radio Astrophysics of the Tata Institute of Fundamental Research. Portions of the work performed at NRL were funded by ONR 6.1 basic research funding. The authors are grateful to the anonymous referee for their insightful comments and constructive suggestions, which greatly improved this work.
\end{acknowledgments}

\clearpage

\appendix
\section{Observation Details}

\startlongtable
\begin{deluxetable}{lccc} \label{table:observations}
\digitalasset
\tablecaption{Details of the start and duration of each observation epoch, along with the number of giant pulses detected.}
\tablehead{\colhead{Date} & \colhead{Start} & \colhead{Duration} & \colhead{$N_{\rm GPs}$} \\ 
\colhead{} & \colhead{MJD} & \colhead{hr} & \colhead{} } 

\startdata
2019-01-13 & 58496.95 & 1.00 & 3 \\
2019-02-04 & 58518.88 & 1.35 & 0 \\
2019-02-25 & 58539.80 & 1.13 & 0 \\
2020-12-06 & 59189.08 & 1.00 & 114 \\
... & ... & ... & ... \\
2024-12-20 & 60665.00 & 0.61 & 0 \\
2025-01-02 & 60677.99 & 1.17 & 7 \\
2025-01-30 & 60705.90 & 0.96 & 0 \\
2025-02-16 & 60722.95 & 0.83 & 0 \\
\enddata
\end{deluxetable}

\section{Giant Pulse Properties}

\startlongtable
\begin{deluxetable}{lccc} \label{table:gpproperties}
\digitalasset
\tablecaption{Properties of the detected giant pulses. The `component' column refers to the component of the average pulse profile with which the GP is phase-aligned; for GPs that do not align with any specific component, the entry is left blank.}

\tablehead{\colhead{Arrival Time} & \colhead{S/N} & \colhead{Width} & \colhead{Component} \\ 
\colhead{MJD} & \colhead{} & \colhead{$\mu$s} & \colhead{}} 

\startdata
58496.95630282 & 9.43 & 143.36 & MP2 \\
58496.96308115 & 7.86 & 92.16 & MP2 \\
58496.97720847 & 11.48 & 10.24 & MP3 \\
58694.41785441 & 7.16 & 20.48 & MP2 \\
... & ... & ... & ... \\
60302.01720319 & 19.70 & 10.24 & MP3 \\
60314.99123644 & 7.43 & 30.72 &  \\
% 60314.99264665 & 9.12 & 10.24 & MP2 \\
% 60329.97455816 & 7.44 & 92.16 & MP2 \\
% 60355.87657323 & 7.56 & 20.48 & MP3 \\
% 60430.72683321 & 9.78 & 10.24 & MP3 \\
% 60430.73356138 & 26.94 & 10.24 & MP3 \\
% 60434.66585921 & 7.05 & 61.44 & MP1 \\
% 60475.56035485 & 7.14 & 61.44 & MP2 \\
% 60505.48852223 & 9.05 & 92.16 & MP2 \\
% 60505.49011955 & 10.45 & 92.16 & MP2 \\
% 60505.49017798 & 7.03 & 143.36 & MP2 \\
% 60505.50019594 & 8.46 & 92.16 & MP2 \\
% 60505.50438852 & 10.86 & 92.16 & MP2 \\
% 60505.50602284 & 10.03 & 143.36 & MP2 \\
% 60505.50607452 & 7.57 & 92.16 & MP2 \\
% 60505.50627920 & 7.05 & 92.16 & MP2 \\
% 60505.50877756 & 8.80 & 143.36 & MP2 \\
% 60505.51122954 & 9.94 & 40.96 & MP1 \\
% 60505.51138628 & 8.34 & 92.16 & MP2 \\
% 60505.51139104 & 9.20 & 61.44 & MP2 \\
% 60505.51205999 & 7.40 & 143.36 & MP2 \\
% 60505.51546613 & 7.55 & 92.16 & MP2 \\
% 60505.51588025 & 7.74 & 61.44 & MP2 \\
% 60505.51589315 & 7.97 & 92.16 & MP2 \\
% 60505.51604400 & 10.87 & 61.44 & MP2 \\
% 60505.52108059 & 7.96 & 61.44 & MP2 \\
% 60505.52163064 & 21.22 & 10.24 & MP3 \\
% 60561.36847335 & 7.09 & 2.56 & MP3 \\
% 60561.37055738 & 7.90 & 2.56 & MP3 \\
% 60561.38685436 & 11.20 & 2.56 & MP3 \\
% 60561.39063057 & 12.38 & 2.56 & MP3 \\
% 60650.06238433 & 8.57 & 1.28 & MP3 \\
% 60677.99638143 & 7.25 & 1.28 & MP3 \\
% 60678.00157529 & 12.66 & 2.56 & MP3 \\
% 60678.00193204 & 7.56 & 2.56 & MP3 \\
% 60678.00528265 & 7.01 & 2.56 & MP3 \\
% 60678.00529149 & 15.12 & 2.56 & MP3 \\
60678.03061684 & 7.11 & 3.84 & MP3 \\
60678.03428082 & 8.00 & 2.56 & MP3 \\
\enddata
\end{deluxetable}

\tablecomments{Tables~\ref{table:observations} and \ref{table:gpproperties} are published in their entirety in the machine-readable format. Portions are shown here for guidance regarding their form and content.}

\bibliography{references}{}
\bibliographystyle{aasjournalv7}

\end{document}